# IMAGINARY Math Exhibition using Low-cost 3D Printers

Marco Rainone, Carlo Fonda and Enrique Canessa

*ICTP Scientific FabLab*
*International Centre for Theoretical Physics, Trieste, Italy*
*e-mail:* **scifablab@ictp.it**

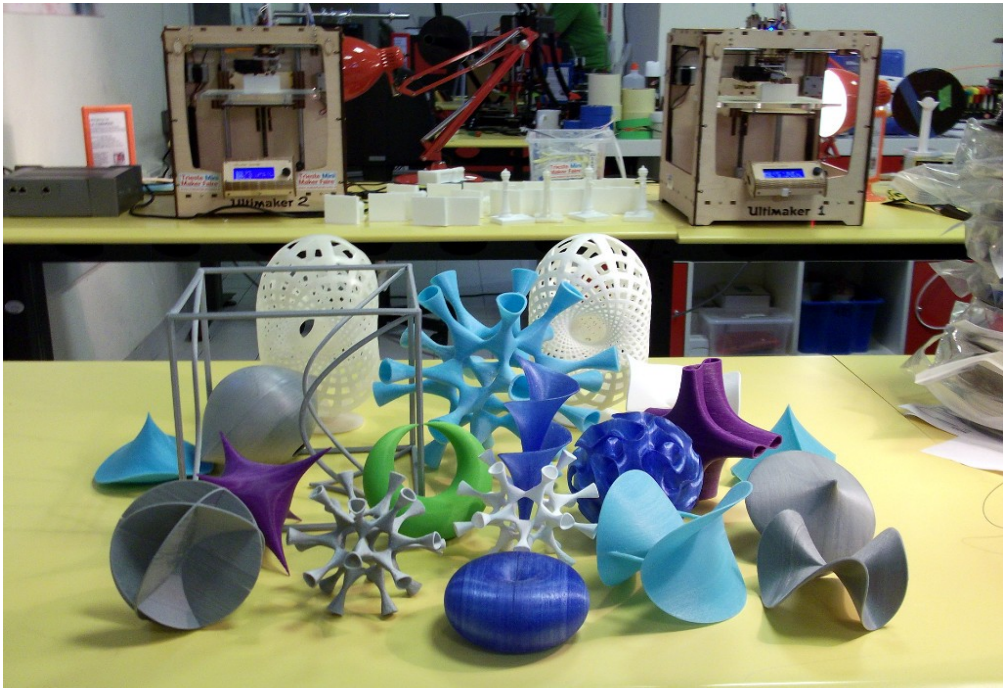

**Overview First Trial**

*We have made an attempt to reproduce 17 objects of the IMAGINARY Open Mathematics Exhibition (www.imaginary.org) using low-cost, desktop 3D printers. The IMAGINARY open math is an international project by the Mathematisches Forschungsinstitut Oberwolfach in Germany and includes galleries of volumetric objects that are unique, have aesthetic appeal and mathematical meaning. We illustrate here the printing of these diverse learning materials using new 3D affordable printing technologies based on Fused Deposition Modelling (FDM) and the use of biodegradable plastic PLA. The final goal is to support museums, schools and higher education institutions in countries with lower scientific infrastructure.*

*ICTP SciFabLab – Sept. 2014*

## Motivation

IMAGINARY started in 2008 as an interactive travelling exhibition that presents visualisations, interactive installations, virtual worlds, 3D objects and their mathematical background in an attractive and understandable way. Over the last years it has developed into an open source organization including a platform (www.imaginary.org) and a network for interactive and participative math communication. The aim of the platform "IMAGINARY –Open Mathematics" is to provide a space for the presentation and development of math exhibitions [1].

The exhibition has visited more than 125 cities in 29 countries so far and there are even some permanent installations for example in the Deutsches Museum in Munich and in the MoMath in New York. The didactical and aesthetic concept inspires the visitors for mathematics and sets out to evoke interest and curiosity for the theoretical background: by offering visuell impressions, live interactions as well as guided tours and individual assistance on site. The target group of the platform includes museums, universities and schools.

All contents of IMAGINARY are made available to a broad audience under a free licence and can thus be reproduced and used for individual exhibitions and events. Moreover, the platform provides an opportunity for everyone interested to contribute with their own material and serves as a hub for exchange of ideas in the field of math communication, a field that has seen many advances in the last years. In this report, we illustrate the possibility of printing 17 different IMAGINARY learning materials using new affordable 3D printing technologies based on Fused Deposition Modelling (FDM). The main author for most of the 3D IMAGINARY data is Oliver Labs, with original data from Herwig Hauser and the FORWISS Institute Passau. The Lawson surface is by the Geometriewerstatt (University of Tübingen).

Low-cost 3D Printing although still in its infancy, is rapidly maturing, with seemingly unlimited potential. With its capability to reproduce 3D objects –from, e.g., medical prostheses up to complex mathematical surfaces– the technology holds a particularly promising future for science, education and sustainable development (see our free book [2]). The 3D printing industry started in the late 1980s (with some initial experiments in the 1970s), but these expensive machines limited the use to professionals. The current expansion of low-cost 3D technologies has benefited from the expired 3D printing patents for FDM, where objects are built up layer by layer with extruded melted plastic. New 3D printing technologies also benefit from the open-source movement (for both open software and open hardware), and from the free sharing of digital files (.STL format) via Internet. We have used wheels of filaments made of biodegradable plastic PLA (Polylactic acid), an environmentally friendly material derived from corn starch. Our final goal is to support museums, schools and higher education institutions in countries with lower scientific infrastructure.

## Tools Used in the Project

*Low-cost 3D Printers:* For printing, we have used two low-cost *Ultimaker* 3D printers, one of which has a double extruder. Compared to the other models available in our Scientific FabLab, these printers have proven to be quite reliable over time after heavy usage. Both models are quite similar and in case of failure it was possible to transfer printing jobs easily from one to another *Ultimaker* with essentially no changes. These printer models use PLA thermoplastic filament of 3mm which is easily available on the market in a range of colors.

***Hardware and Software:*** For processing, we used a Laptop with these features: Processor Intel Celeron 1.6 Ghz, 64bit, Hard Drive 1Tb, 8 Gb Ram, Operating system Windows 7 Home Premium.

For 3D printing there are a number of tools −both free or proprietary, that allow the work required to get an object printed. However, some specific options for printing that may be present in one software may not be available in other, or another software may be more efficient in the execution of some operations. The situation can also change with updated versions of the same type of software. It is therefore difficult to determine a-priori which is the most suitable toolchain to develop objects for 3D printing at low-cost. One often relies on the experience gained and on the efficiency and ease of use of one particular machine. A number of programs used in the project is given next.

> **Autodesk Meshmixer** *Ver.10.5.79*, 64 bit (www.meshmixer.com): free prototype design tool from Autodesk, based on high-resolution dynamic triangle meshes. We used it because it is particularly effective when displaying STL files before proceeding with any other operations performed by other software.
>
> **Netfabb Studio** (www.netfabb.com): Proprietary software which provides a free version called Netfabb Basic. In addition to various image manipulation and editing of three-dimensional objects, with this software we fixed problems in the geometry of the construction of a 3D model before sending it to print. Given that the 3D printouts are usually slow, this verification allows to save time and money. For the processing executed in the project we used "**Netfabb Studio basic for Ultimaker**", a particular version of Netfabb downloaded from the website *Ultimaker*, which has a few more options than the free version Netfabb Basic -see Netfabb Studio basic for Ultimaker i386 ( http://software.ultimaker.com/).
>
> **OpenSCAD** *Ver.2013.06* (www.openscad.org): free software for creating solid 3D CAD objects. This software is a "compiler" that reads instructions from a script and renders a 3D model out of it. We used OpenSCAD to perform some operations, such as merge objects and get the STL 3D file to be sent to print.
>
> **Ultimaker Cura** *Ver.14.03* (http://software.ultimaker.com/): software provided by the *Ultimaker* for its printers and can also be used for printers like RepRap. It contains a 3D model preview, a Slicer, GCode generator, and GCode sender to the printer. Before slicing one can also perform operations directly on the 3D model, such as rotation, change size, etc.
>
> **Software for Converting 3D Formats:** Before processing the figures, the first problem found was to convert into STL format the files of some of the IMAGINARY three-dimensional objects given in X3D format. This is so because, e.g., the "Cura" program is not currently able to import them. For this operation, we tested two tools: Meshlab (open-source, developed by the ISTI–CNR research center, http://meshlab.sourceforge.net/) and Netfabb (www.netfabb.com).

## **Printing Outputs: IMAGINARY Sculptures**

**The actual printings shown below are the product of multiple trials of small models or parts of the IMAGINARY complex objects by applying procedures that were refined gradually.** Only two figures were printed directly with "Cura", applying magnifications or rotations prior to the actual print of the object. All other math objects were processed or manipulated with tools such as Meshlab before slicing through "Cura". A winning strategy was to group the math objects according to sequences of similar operations that could be applied on them.

We briefly list below the operations performed on 17 math objects. The necessary steps for printing each object will be described in a second report to be available at http://scifablab.ictp.it

1 **Helix_200mm_th3p0mm**: *directly manipulated and printed with "Cura"*

A visual analysis of this object highlighted characteristics such as to make it printable through "Cura". Once enlarged, it had a base wide enough that it can be printed without further operations.

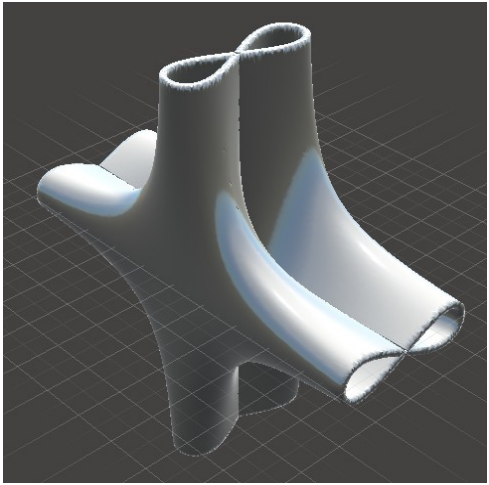
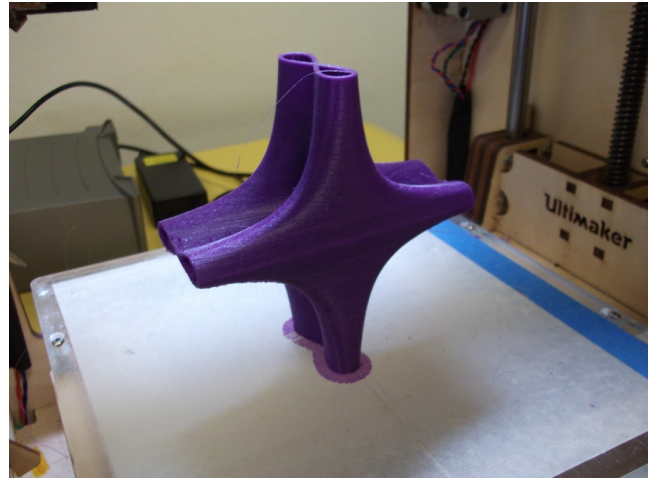

2 **Lawson_201mm_th2p1mm**: *written an OpenSCAD script to add a support*

This figure was first enlarged and then it was joined to a support base through an OpenSCAD script.

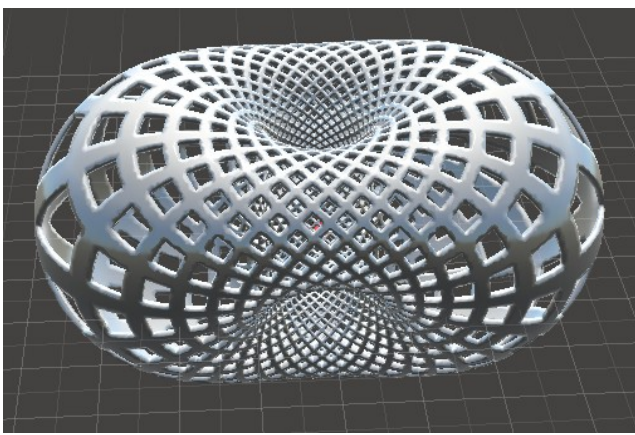
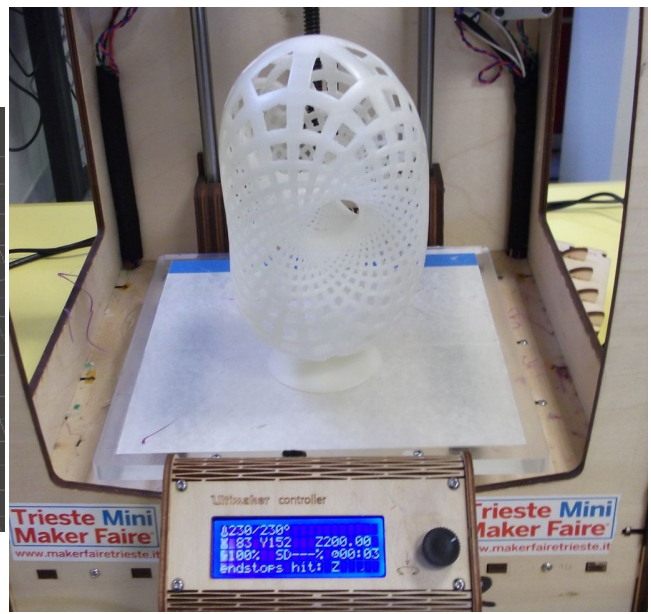

# Objects divided in two halves

For the following objects we have carried out similar operations using Netfabb:

- The 3D model was eventually corrected.
- If necessary, the object was possibly reduced or enlarged to optimize printing, and
- The object has been divided into two halves, which were printed together or separately through "Cura" Once finished and cleaned, the two parts were glued using cyanoacrylate.

### 3    Barth_65_50fin_206mm_th2p1mm

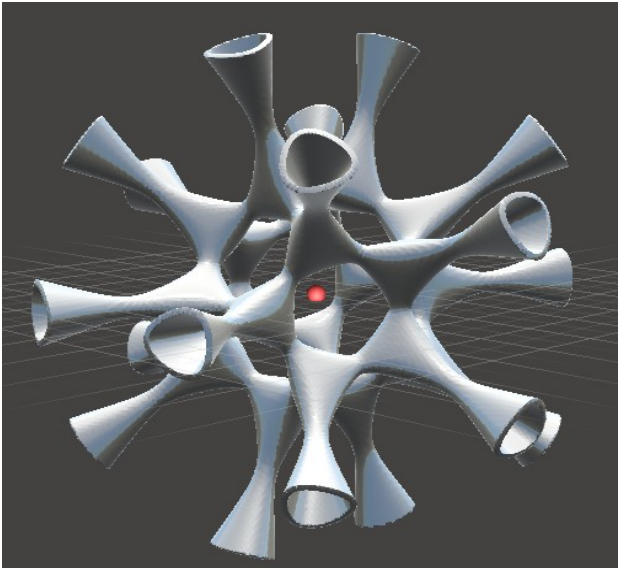
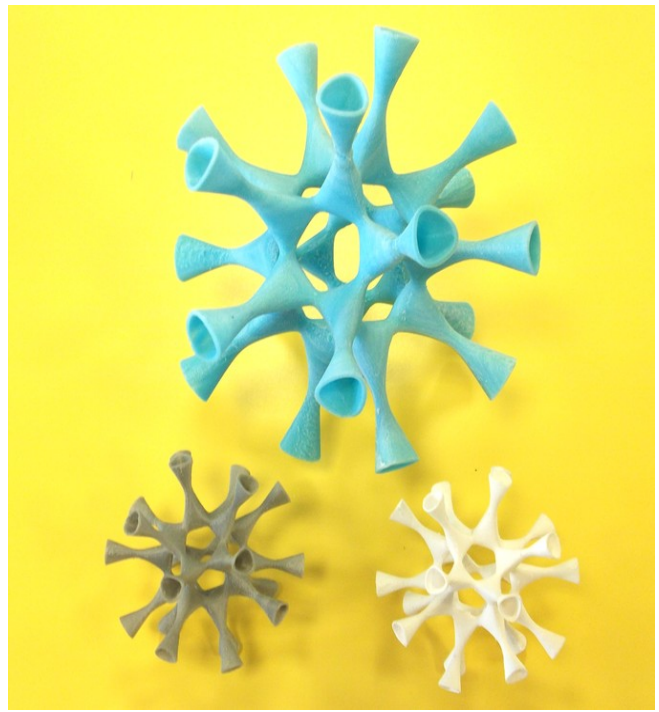

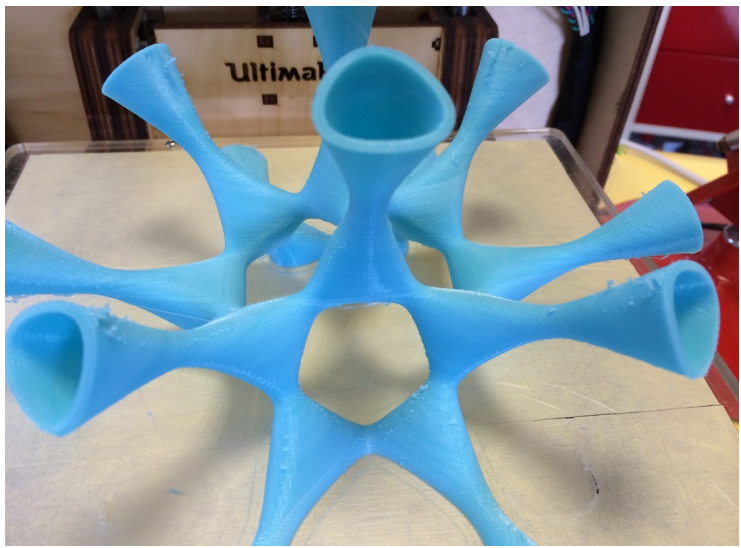

## 4 Calypso_with_support 200mm_th2mm

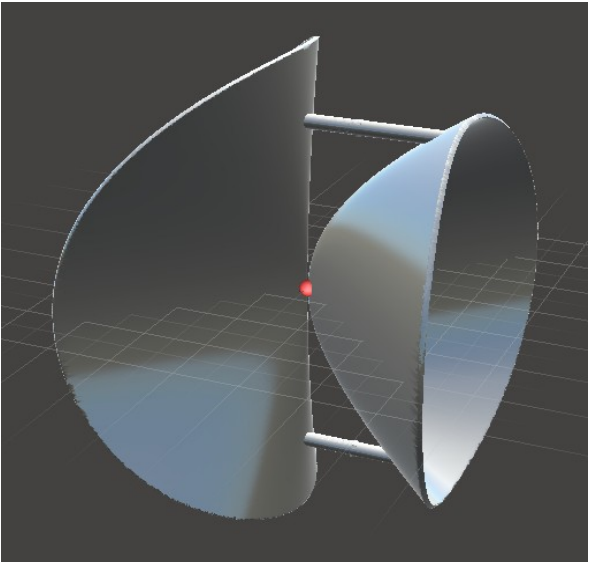
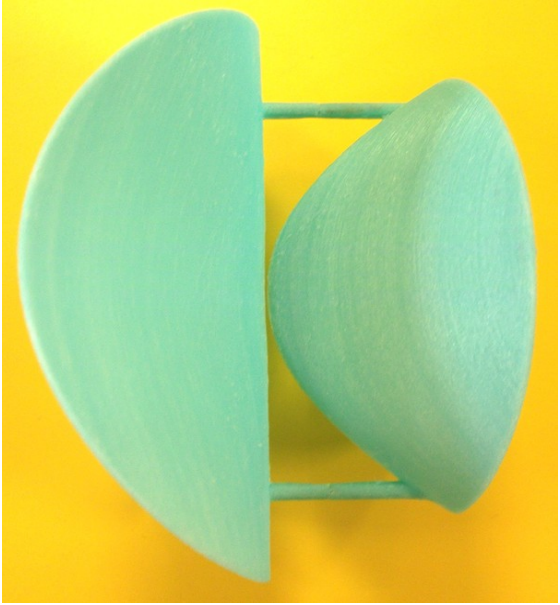

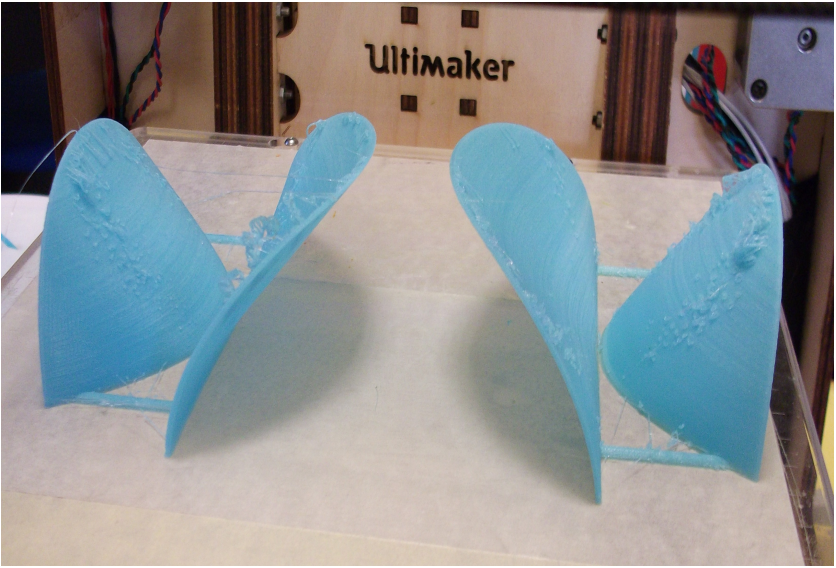

## 5    **DiniSurface_299mm_th2p1mm**

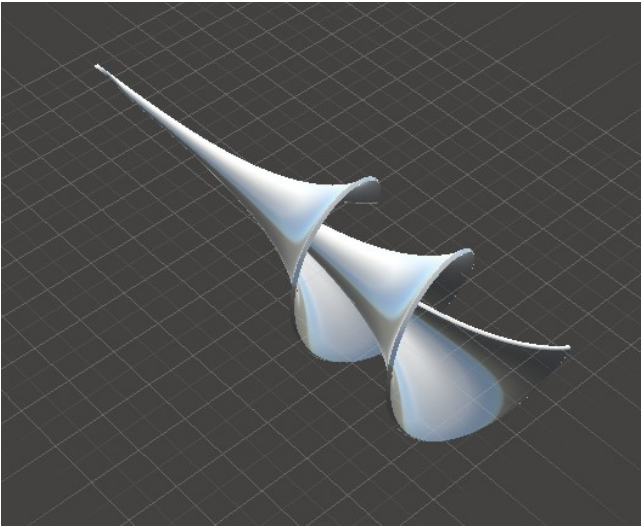
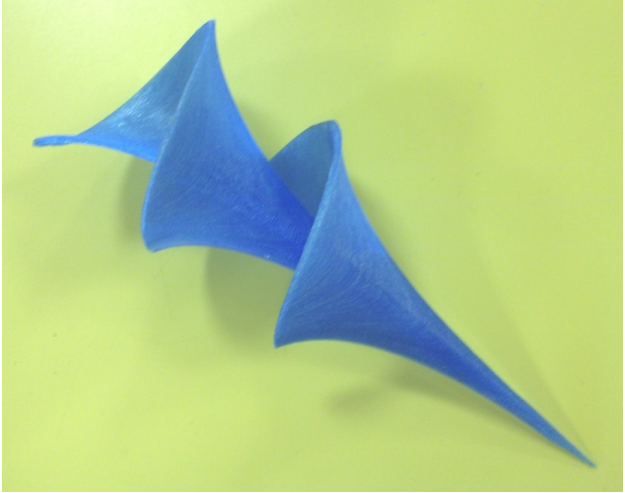

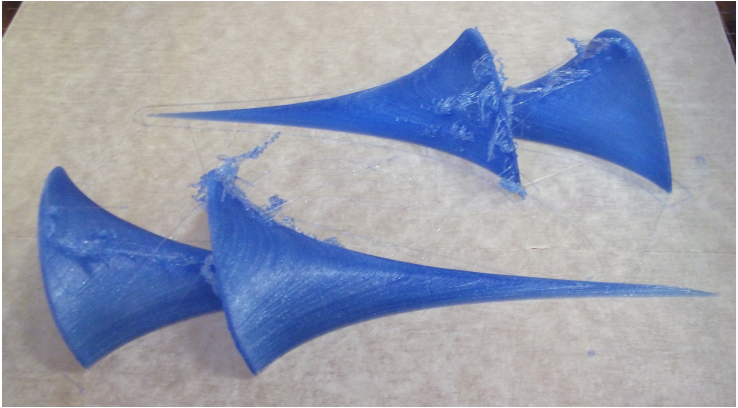

## 6    **Spitz_223mm_th2p5mm**

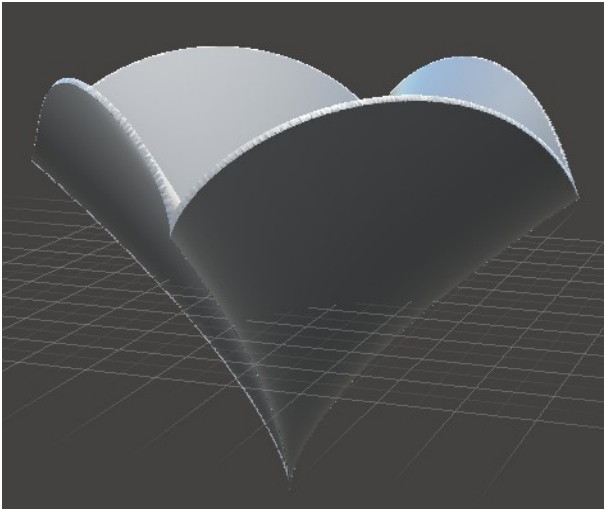
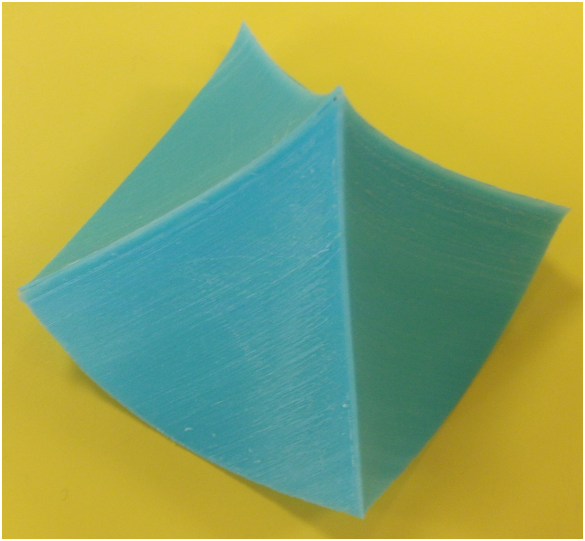

## 7 **Kreisel**_hollow_200mm_th3p1mm

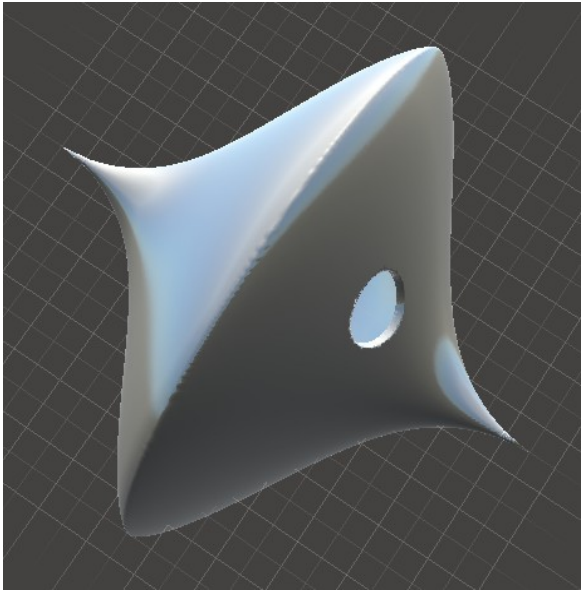
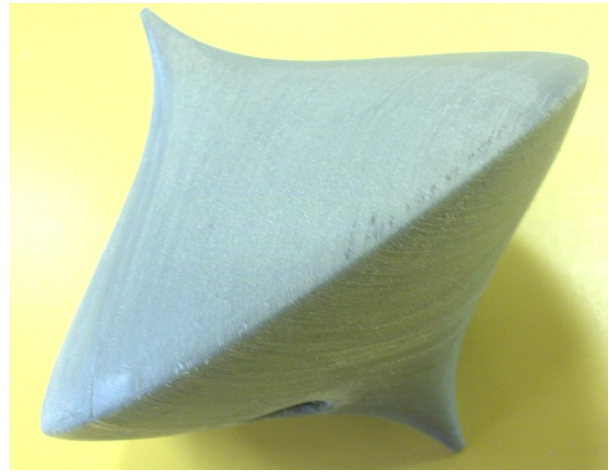

## 8 **Distel**_200mm_full

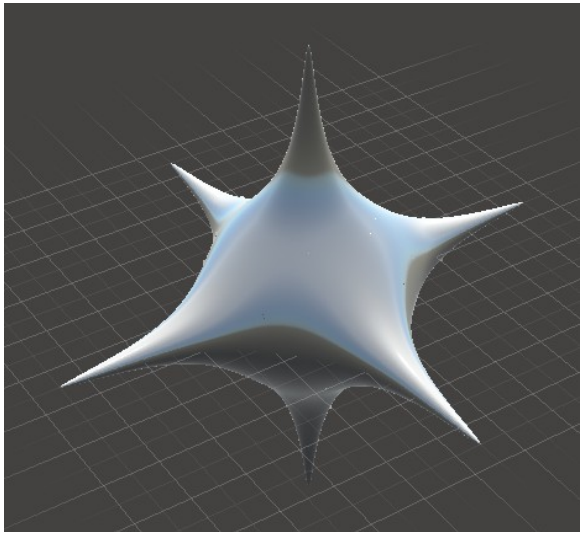
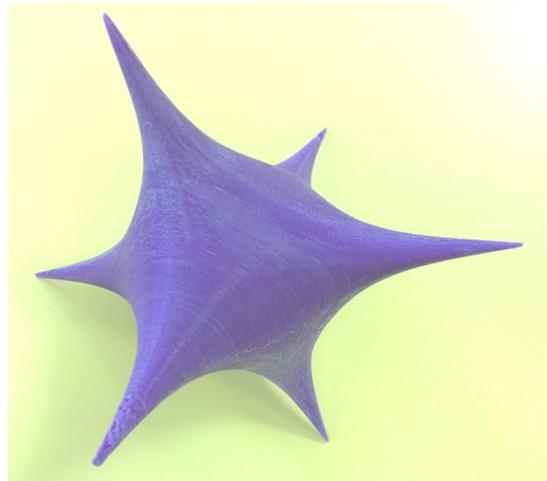

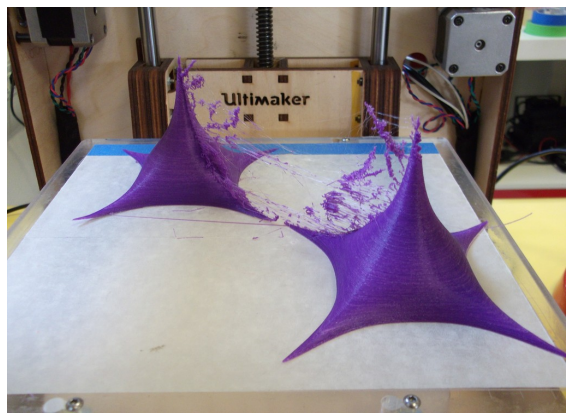

## 9 Lemon_offset_215mm

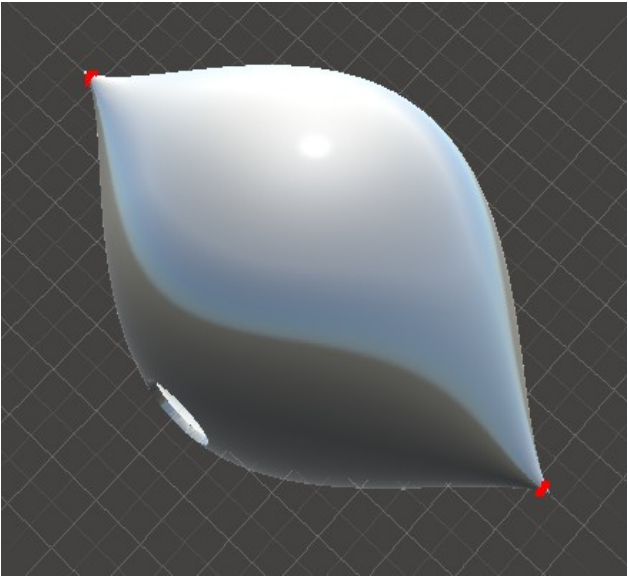 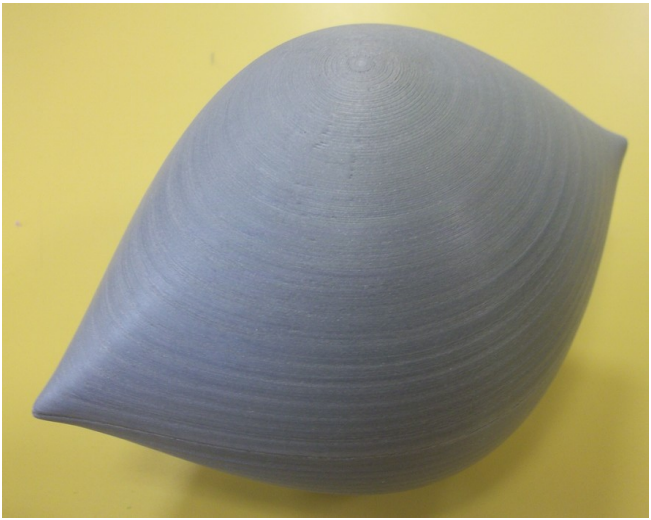

## 10 Nepali_empty_200mm_th3p0mm

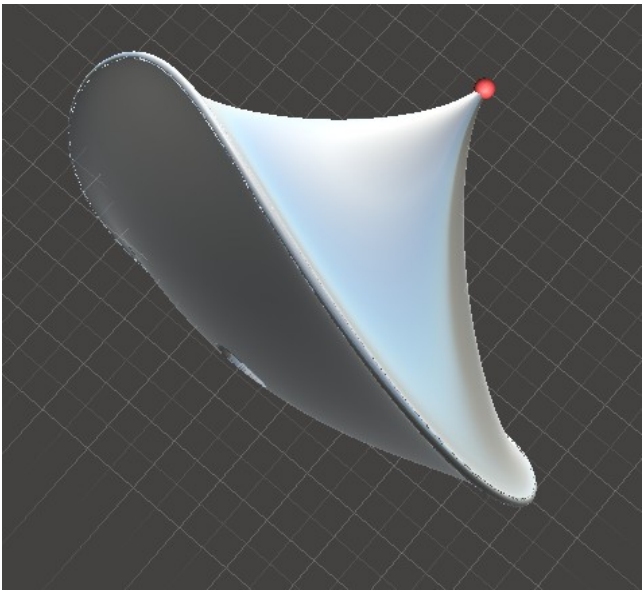 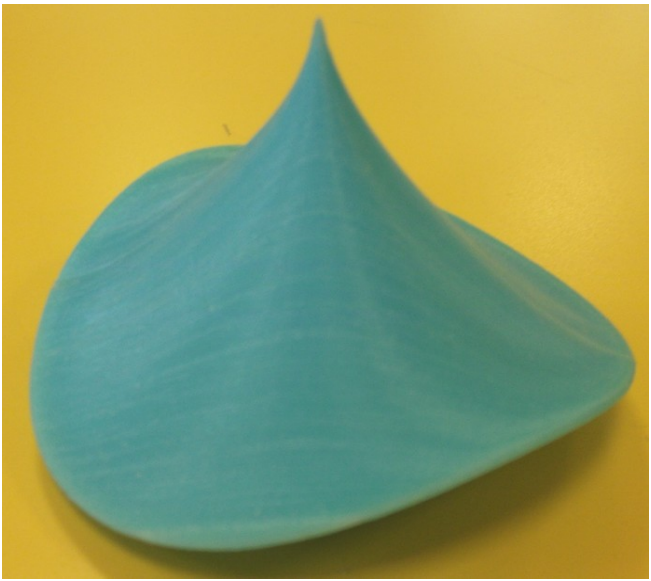

## 11  Schneeflocke_200mm_th3p5mm

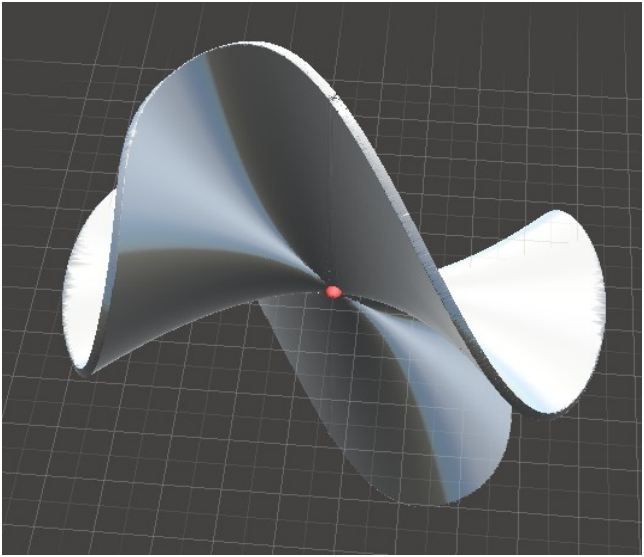
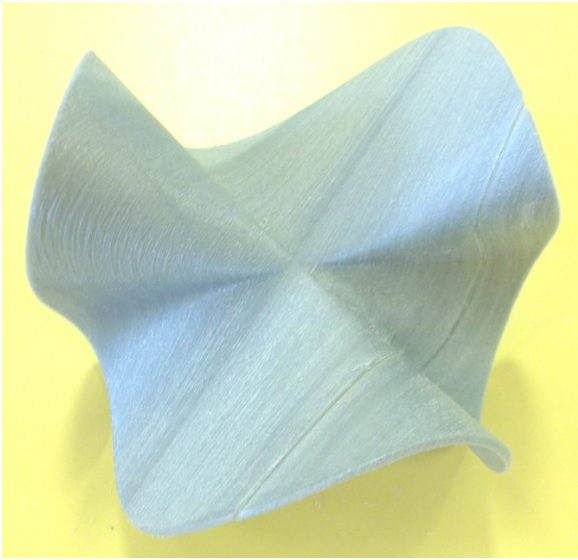

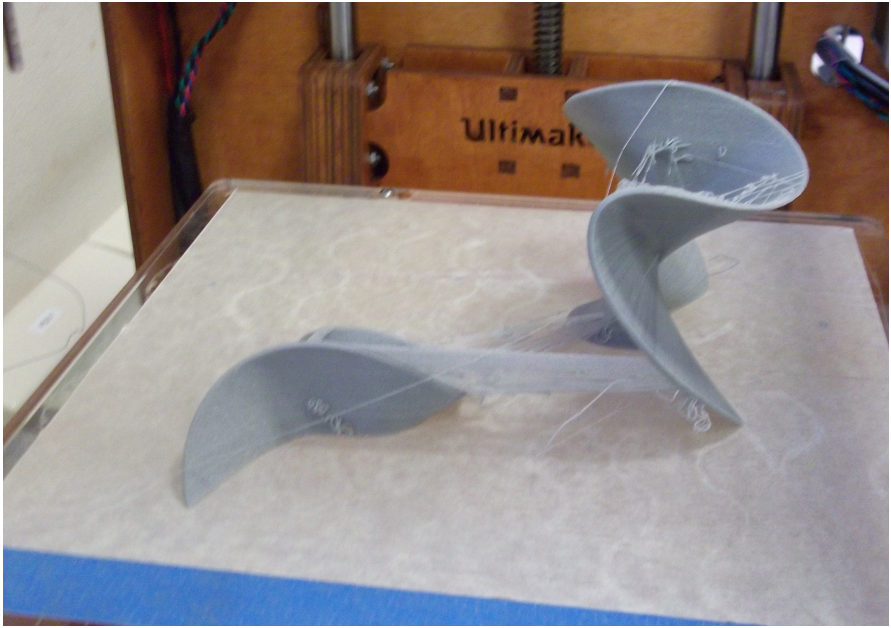

## 12  Dullo_200mm_th3p1

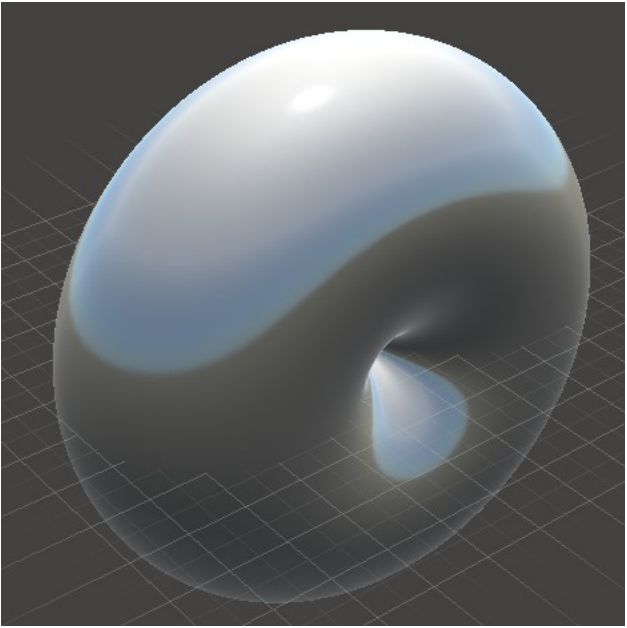
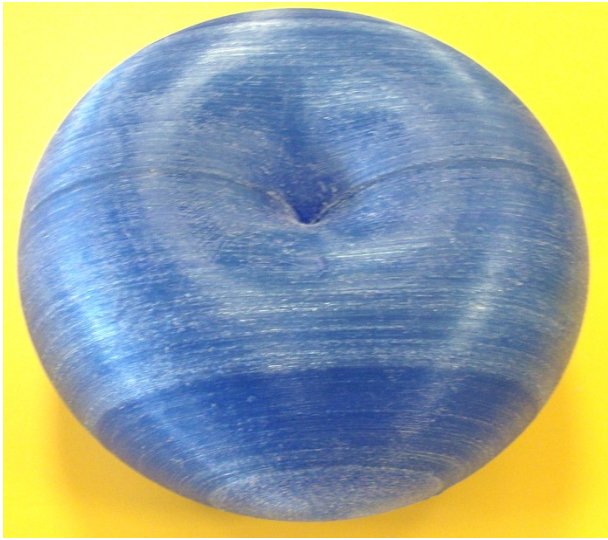

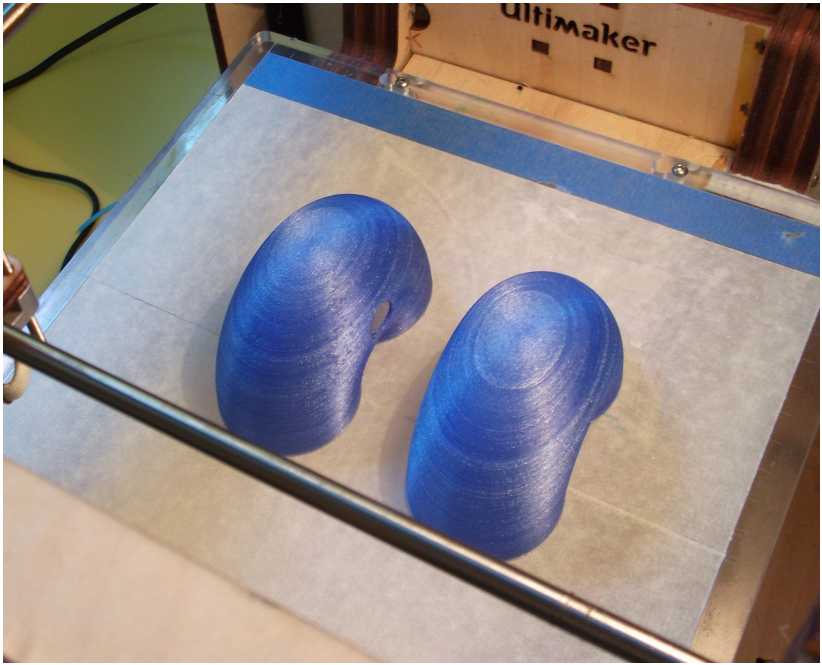

## 13   Gyroid  199mm  th3p0mm

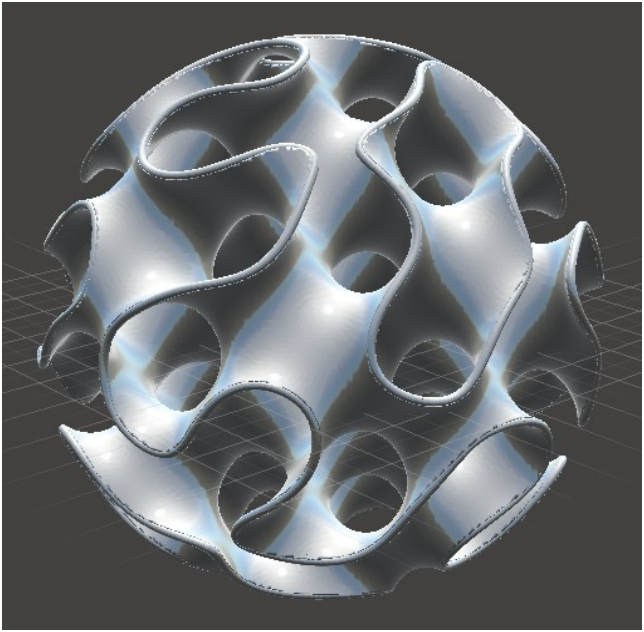
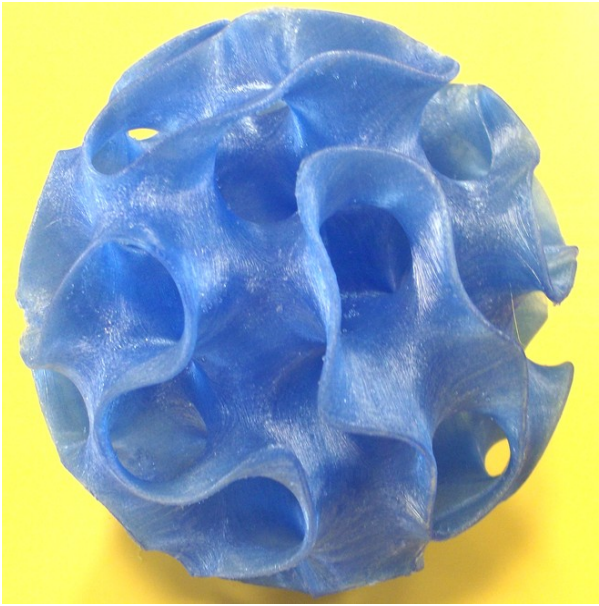

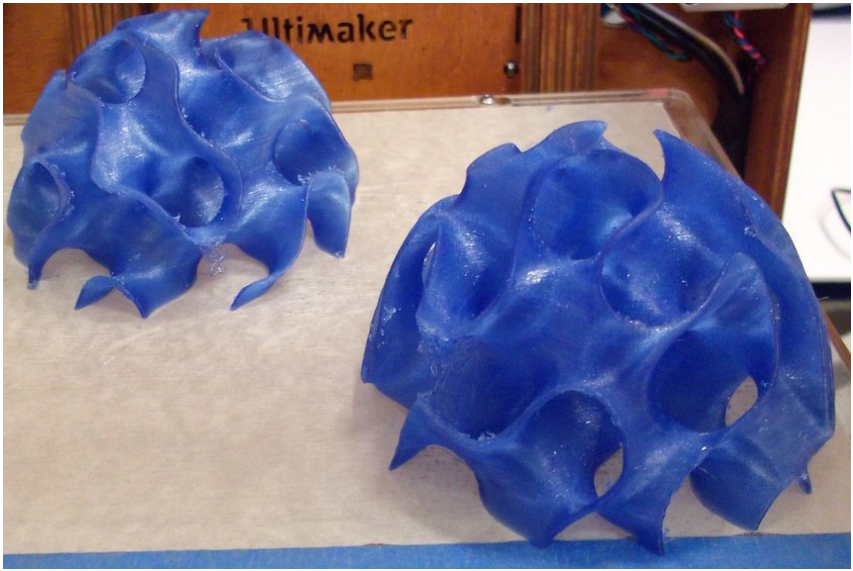

## 14   Croissant_empty_200mm_th3p0mm

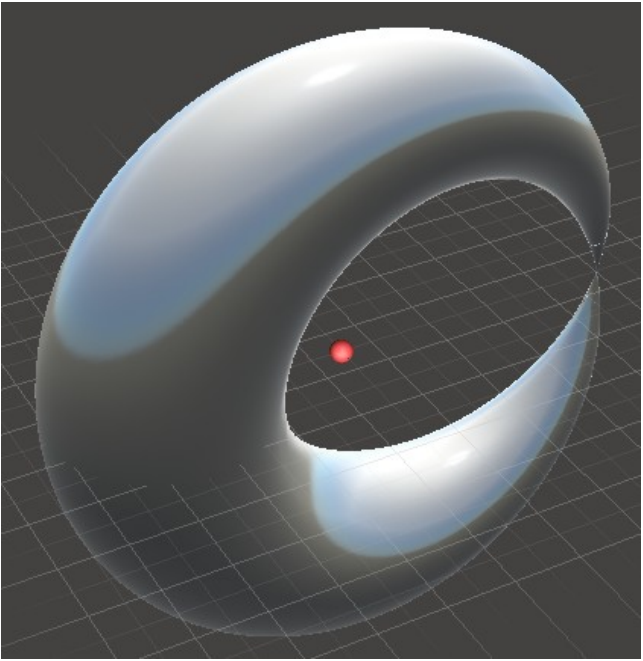
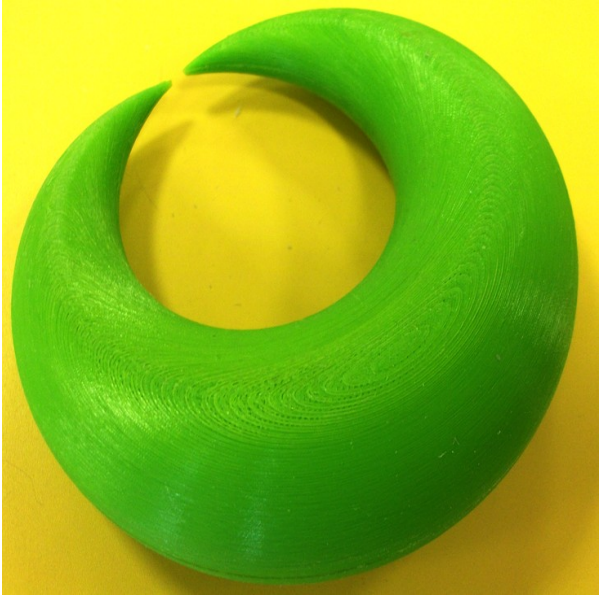

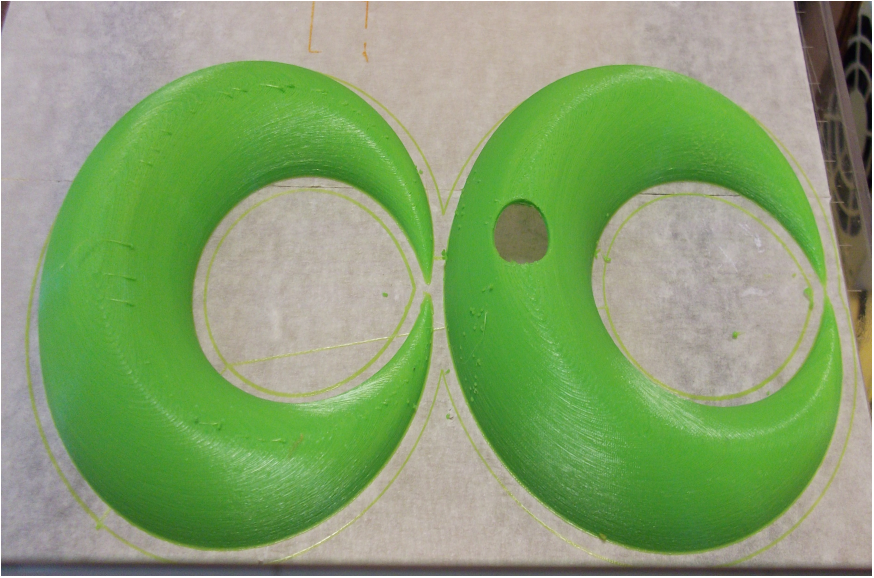

## 15   <u>**Tuelle**  200mm  th3p0mm</u>

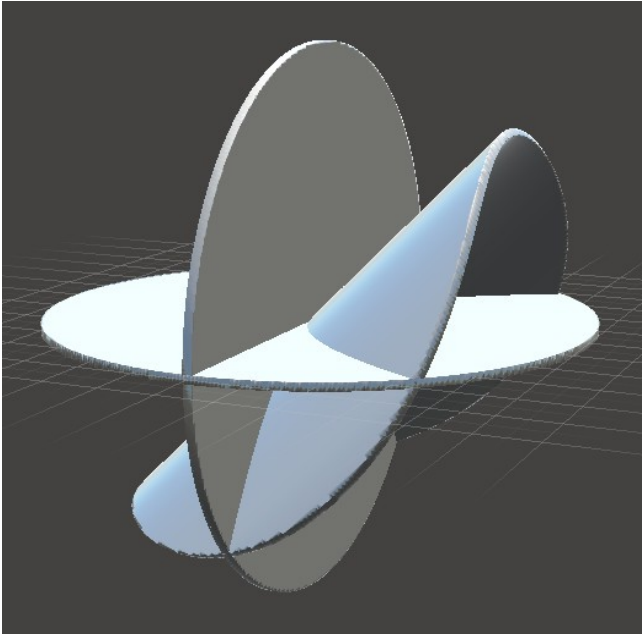 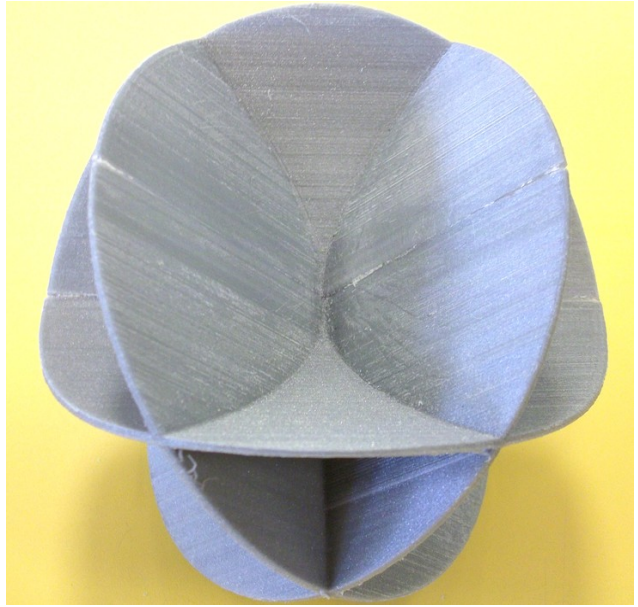

## 16   <u>**Visavis**  211mm  th4p0mm</u>

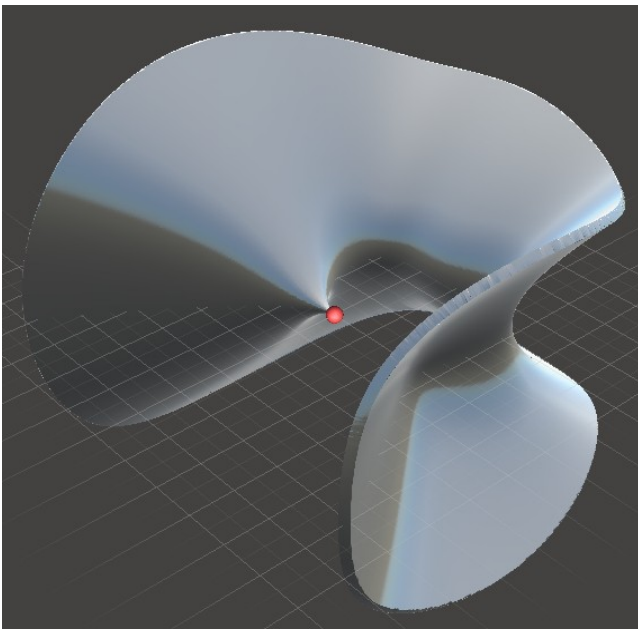 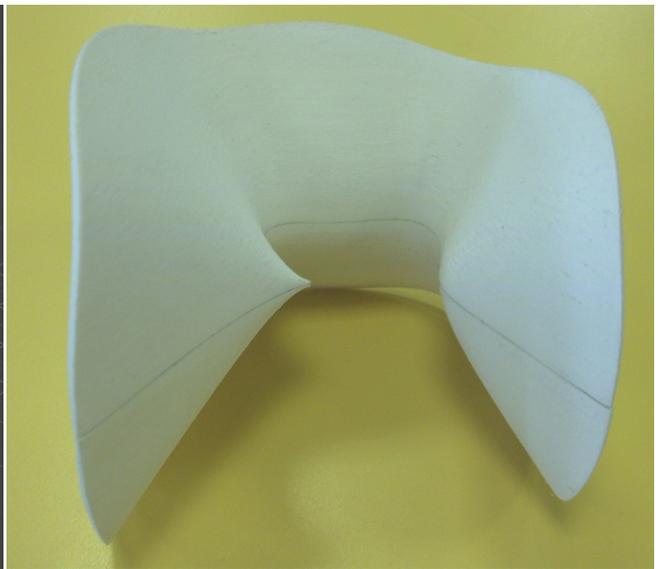

**17  <u>SpaceCurveInCube_206mm_th5p0mm:</u>** *object divided in 5 parts*

Despite appearances, this object of the IMAGINARY exhibit was the most laborious object to accomplish. We proceeded with NetFabb to divide the object into 5 parts (4 parts of the outer faces of the cube plus the inner element), which were then assembled by gluing it with cyanoacrylate.

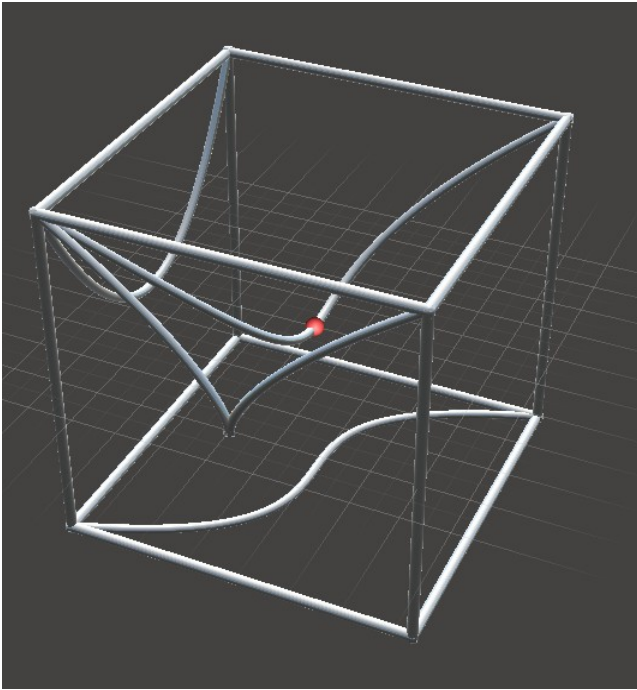
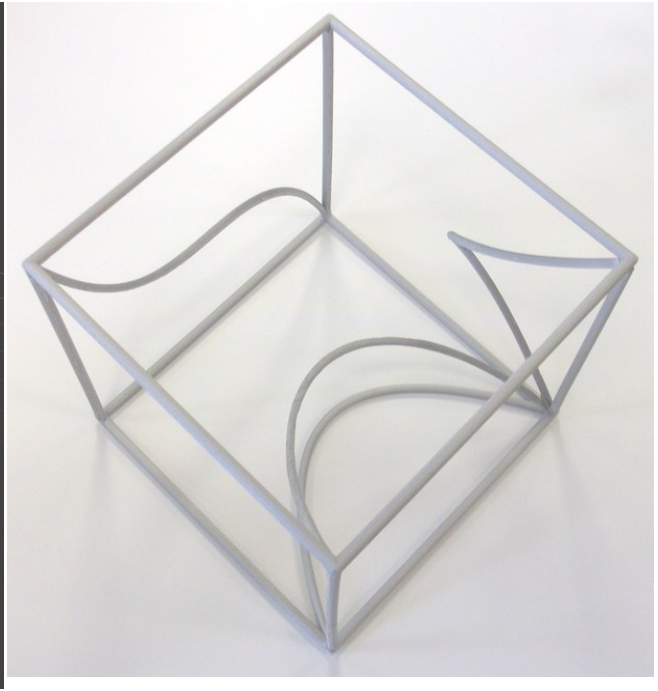
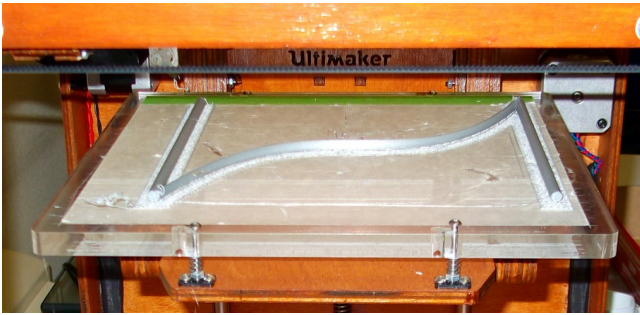
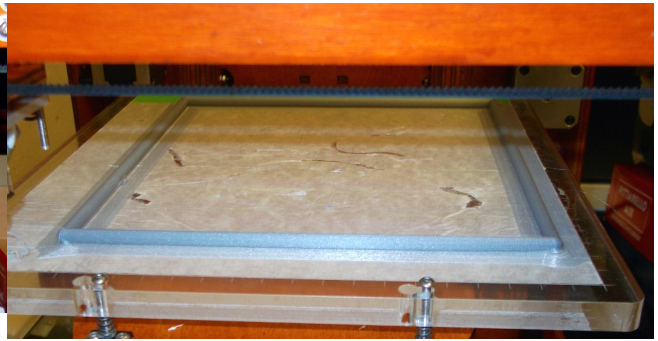
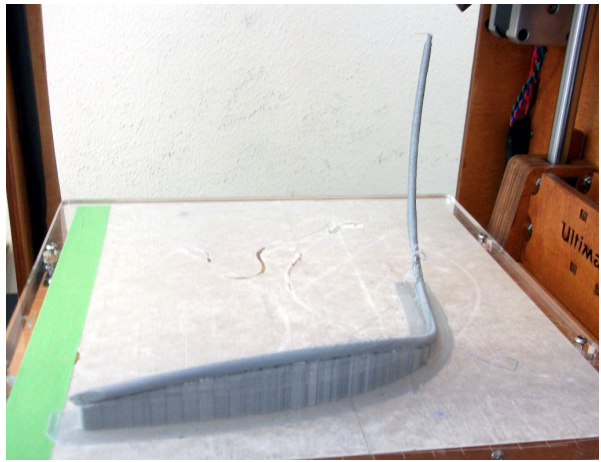

# Estimated Costs

In our estimations we include the length of the filament used and the parameters that "Cura" provides at the end of its process of the slicing. For the calculation, the density ρ of PLA is 1240 kg/m$^3$  The filament is provided in coils with these characteristics given by the manufacturer:

| Coil parameters | Value | Units | Symbol |
|---|---|---|---|
| Reel mass | 1,00 | *Kg* | m |
| Reel costs | 25,00 | *Euro* | C |
| Filament Diameter | 3,00 | *mm* | d |

The lenght of reel is calculated by the relation:  L = m / [ ρ * π * (d/2)²].   For the cost per meter of filament, we divide the cost C of the coil by the number L of calculated meters.

| Plastic | Lenght (m) | Filament cost per meter (Euro/m) |
|---|---|---|
| PLA | 114,0895649437 | 0,2191260876 |

The results for the costs of material PLA used for each object are given in the Table below:

**The respective mathematical equations, meaning and characteristics of these beautiful algebraic surfaces can be found in the IMAGINARY Open Mathematics website: www.imaginary.org**  In particular, the Barths sextic surface of degree 6 (sextic) has 65 singularities when also counting the 15 invisible ones which are infinitely far away. The exact equation of Barth's sextic is $P_6 − αK^2 = 0$, where $P_6$ = ( $τ^2x^2−y^2$)( $τ^2y^2−z^2$)( $τ^2z^2− x^2$), τ = 1/2(1+√5) is the golden ratio, α = 1/4  (2τ+1)=1/4 (2+√5) and K = $x^2+y^2+z^2 − 1$ describes a sphere of radius 1. There are also poster sets constaining information about objects in the IMAGINARY exhibitions.

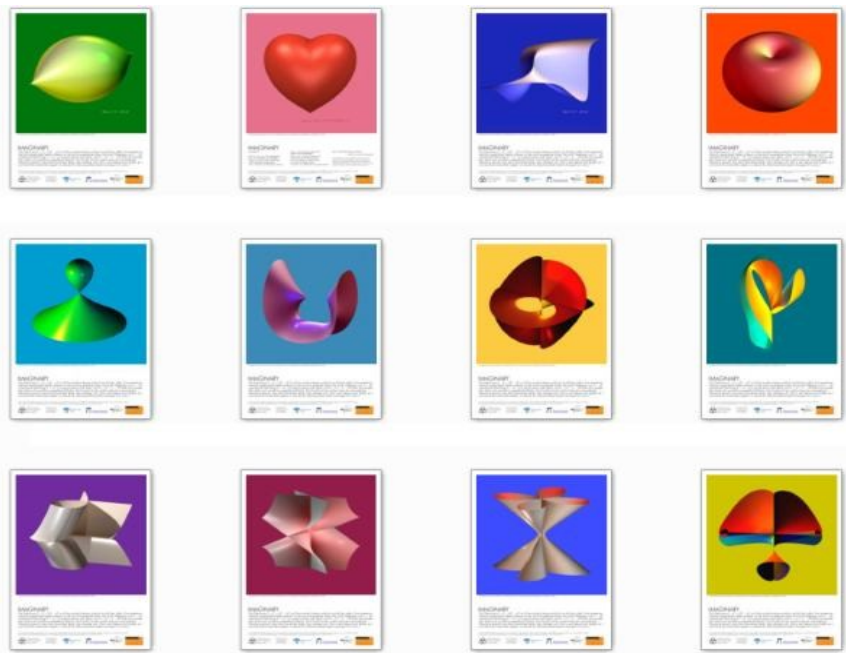

|   | **FileName Object IMAGINARY Exhibition** | **Author** | **Size of the 3D Printing (mm)** | **Length of Filament Used (m)** | **Cost Object** |
|---|---|---|---|---|---|
| 1 | **Barth**_65_50fin_206mm_th2p1mm, Ver. 1 | *Oliver Labs* | 100 x 100 x 100 | 2,56 | **0,56 €** |
| 1(a) | **Barth**_65_50fin_206mm_th2p1mm, Ver. 2 | *Oliver Labs* | 200 x 200 x 200 | 19,52 | **4,28 €** |
| 2 | **Calypso**_with_support_200mm_th2mm | *FORWISS, Herwig Hauser, Oliver Labs* | 86 x 120 x 84 | 3,77 | **0,83 €** |
| 3 | **Croissant**_empty_200mm_th3p0mm | *FORWISS, Herwig Hauser, Oliver Labs* | 95 x 100 x 35 | 3,50 | **0,77 €** |
| 4 | **DiniSurface**_299mm_th2p1mm | *Oliver Labs* | 180 x 50 x 50 | 1,69 | **0,37 €** |
| 5 | **Distel**_200mm_full | *FORWISS, Herwig Hauser, Oliver Labs* | 98 x 98 x 98 | 4,25 | **0,93 €** |
| 6 | **Dullo**_200mm_th3p1 | *FORWISS, Herwig Hauser, Oliver Labs* | 90 x 90 x 45 | 4,14 | **0,91 €** |
| 7 | **Gyroid**_199mm_th3p0mm | *Oliver Labs* | 100 x 100 x 100 | 8,54 | **1,87 €** |
| 8 | **Helix**_200mm_th3p0mm | *Oliver Labs, Herwig Hauser* | 48 x 120 x 120 | 6,48 | **1,42 €** |
| 9 | **Kreisel**_hollow_200mm_th3p1mm | *Oliver Labs, Herwig Hauser* | 100 x100 x 100 | 4,02 | **0,88 €** |
| 10 | **Lawson**_201mm_th2p1mm | *Geometriewerstatt (Univ. of Tübingen)* | 114 x 112 x 192 | 13,23 | **2,90 €** |
| 11 | **Lemon**_offset_215mm | *Herwig Hauser, Oliver Labs* | 95 x 95 x 160 | 8,37 | **1,83 €** |
| 12 | **Nepali**_empty_200mm_th3p0mm | *FORWISS, Herwig Hauser, Oliver Labs* | 96 x 96 x 62 | 3,57 | **0,78 €** |
| 13 | **Schneeflocke**_200mm_th3p5mm | *Oliver Labs, Herwig Hauser* | 70 x 100 x 100 | 2,92 | **0,64 €** |
| 14 | **SpaceCurveInCube**_206mm_th5p0mm | *Oliver Labs* | 180 x 180 x 180 | 10,89 | **2,39 €** |
| 15 | **Spitz**_223mm_th2p5mm.stl | *FORWISS, Herwig Hauser* | 100 x 76 x 100 | 1,97 | **0,43 €** |
| 16 | **Tuelle**_200mm_th3p0mm | *Oliver Labs, Herwig Hauser* | 100 x 100 x 100 | 5,21 | **1,14 €** |
| 17 | **Visavis**_211mm_th4p0mm | *FORWISS, Herwig Hauser, Oliver Labs* | 105 x 95 x 100 | 5,34 | **1,17 €** |

## **Remarks**

One of the main limitations for the alternative use of FDM printing technologies in the reproduction of math exhibits, is the reduced dimension available for printing (typically 20x20x20cm) due to the low-cost for the equipment [2]. However, bigger 3D objects can be formed by assembling many small plastic parts as Lego bricks. Another constraint could be the printing quality due to the non-trivial math singularities and form of the complex objects as those in the IMAGINARY and shown here.

We have customized the math objects and tried to keep to the minimum, or even not to use, any extra (thin) support or special spongy structure of plastic built from below to hold the parts of the object that wouldn't be printed otherwise. We have build up and finished the plastic structures by assembling by hand their component parts which have been splitted up to the minimum possible. A great benefit of using low-cost 3D printers is that this entire process produces much less waste than traditional 3D manufacturing, where large amounts of material are trimmed away from the usable part, and that the financial costs and work needed to set up a math exhibition are considerably lower. The final look and touch of the 3D math printings of the beautiful IMAGINARY objects with singularities using this cutting-edge technology are found appealing [1]. These satisfy our initial expectations to inspire curiosity, incentivate deeper understanding and put learning literally in the hands of scholars and new generations of scientists and mathematicians (alternative approaches can be seen in [3-4]).

We hope this preliminary first attempt to reproduce the IMAGINARY Open Mathematics Exhibition using low-cost 3D printers can be refined in the future. Each of the figures has its own unique characteristics of form and shape and therefore it is difficult to identify a common methodology for printing. We share the preliminary results of this first trial to expand the outreach of the "IMAGINARY –Open Mathematics" communication and support mathematics education. We have carried out this study using the ICTP Scientific FabLab's low-cost 3D printing facilities.

Future common objectives to be defined include how to optimize the consumption of plastic filament and provide useful information such as time and material consumed for printing, parameters and settings used for the printers, etc. Work along these lines is in progress and will be published at http://scifablab.ictp.it

*About the ICTP Scientific FabLab*

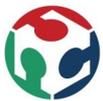
The Science Dissemination Unit (SDU) of the Abdus Salam International Centre for Theoretical Physics (ICTP) in Trieste, Italy has established on August 12, 2014 a Scientific FabLab ("*fabrication laboratory*") infrastructure devoted to creativity and research with special focus on possible applications of benefit for the society. FabLabs are a global network of local labs (www.fablabs.io), enabling invention by providing access to tools for rapid digital fabrication. In particular, the ICTP Scientific FabLab within ICTP campus is a workshop area for Scientists and Makers campus and it offers the possibility of digital fabrication and rapid prototyping for projects in the fields of science, education and sustainable development.

*Acknowledgements*

We would like to thank Prof. Fernando Rodriguez Villegas, Head of the ICTP Mathematics Group, and Prof. Gert-Martin Greuel (Scientific Advisor) and Dr. Andreas Daniel Matt (Project Management, Communication, Media) from the IMAGINARY project by the Mathematisches Forschungsinstitut Oberwolfach.